\documentclass[prb,
superscriptaddress,showpacs,amsmath,amssymb]{revtex4}

\begin{document}

\title{ Andreev hidden spinor coordinates: Standard Model, gravity and $^3$He}

\author{G.E.~Volovik}
\affiliation{Low Temperature Laboratory, Aalto University,  P.O. Box 15100, FI-00076 Aalto, Finland}
\affiliation{Landau Institute for Theoretical Physics, acad. Semyonov av., 1a, 142432,
Chernogolovka, Russia}

\begin{abstract}
This paper is prepared for a special memorial issue of J. Low Temp. Phys. dedicated to memory of Alexander Andreev. I discuss his ideas devoted to the fundamental problem in modern physics -- the origin and validity of the superselection rule, which forbids the superposition of a fermion and a boson state. Andreev suggested extra spin degrees of freedom, due to which the $2\pi$ rotation does not change the sign of the fermionic wave function, and as a result the fermion-boson transmutations becomes possible. Although his approach looks somewhat contradictory from the point of view of the present physics, in principle his ideas can be realized on the more fundamental trans-Planckian level leading to new physics. Different scenarios of the extension of the internal spinor space are considered.
\end{abstract}

\maketitle

\tableofcontents

\section{Introduction. Andreev and spinor coordinate}
\label{SecIntroduction}

A.F. Andreev wrote several papers \cite{Andreev1998,Andreev2000a,Andreev2000b,Andreev2001} devoted to one of the most puzzling problems in modern physics -- the conjecture of the superselection rule, which forbids the linear superpositions of states with even and odd numbers of fermions.
This superposition is incompatible with the Lorentz invariance, since the $O(2\pi)$ transformation changes the sign of the fermion wave function but does not change the sign of the boson wave function. This leads to the superselection rule.\cite{WWW1952} 
According to Andreev, the superselection rule is not self-consistent. Instead he suggested that the spacetime must be extended.  In addition to the $x, y, z, t$ coordinates there should be a special spinor coordinate, which characterizes the internal spinor symmetry on the fundamental level. As a result, the generalized $O(2\pi)$ transformation does not change change the sign of the fermionic wave function and the fermion-boson transmutation is allowed.

Although the particular scenario proposed by Andreev is not consistent with the usual formulation of physics as we know it, his attempts to construct a consistent theory of the fundamental space-time, which may allow for linear superpositions of bosons and fermions, deserve attention and further development. It is not excluded that on a more fundamental level, say, at the trans-Planckian energy scale, the principles of the relativistic quantum field theory can be violated and the boson-fermion transmutation becomes possible. Then the extension of the coordinate space to include the extra spinor degrees of freedom can be the right direction towards new physics. Anyway, the idea of the extension of the internal (spin) space, even if it does not destroy the superselection rule, leads to interesting consequences for superfluid $^3$He. Maybe it will also lead to new physics on a more fundamental level since $^3$He can be used as a platform for the simulation of many new directions in physics.\cite{Volovik2020a,Volovik2023}

Here we consider several examples of the extension of the internal spinor symmetry. This in particular includes the extension of the internal  Lorentz group from $SO(1,3)$ to $SO(1, 4)$, which is discussed for the extension of the Standard Model,\cite{Maiezza2022} see Sections  \ref{SpinorSM} and \ref{Rectangular}. The corresponding relativistic physics, which emerges in the vicinity of the Dirac points in the spectrum of the planar phase of superfluid $^3$He,\cite{Volovik2020,Volovik2022b} is discussed in Sections  \ref{SpinorPlanar} and \ref{VielbeinPlanar}. 
The superposition of states with different numbers of particles suggested by Andreev and the birth of time crystals are described in Section \ref{TimeCrystal}.
 The fermion-boson transmutation due to quantum anomalies and the 1+1 quantum field theory, which is based on the Andreev proposal, are discussed in Section \ref{Anomaly}.
The extension of the Standard Model to lattice theory with several Weyl and Dirac points also suggests the possibility of observing processes connected with the creation or annihilation of single fermions (see Sec. \ref{AndreevWeyl}).

\section{Planar phase of superfluid $^3$He}
\label{PlanarPhase}

Let us start with the superfluid $^3$He, where the so-called planar phase demonstrates the mechanism of the extension of the internal spin space. The discrete $Z_2$ symmetry becomes the continuous $U(1)$ symmetry, which leads to the additional spinor coordinate.

The spin triplet $p$-wave pairing in superfluid $^3$He has the following $2\times 2$ matrix of the gap function:
\begin{equation}
 \hat{\Delta}=  A_{\alpha}^i p_i\sigma^\alpha i\sigma_y \,,
\label{SpinTriplet}
\end{equation}
where $\sigma^\alpha$ are the Pauli spin matrices and $A_{\alpha}^i$  is the $3\times 3$ complex matrix of the order parameter, see the book \cite{Vollhardt1990}.
In  the planar phase of superfluid $^3$He the particular choice of the order parameter is:
\begin{eqnarray}
A_{\alpha}^i=c_\perp \left(\delta_\alpha^i - {\hat z}_\alpha  {\hat z}^i \right) \,,
\label{OP1}
\end{eqnarray}
where $c_\perp$ is the "speed of light" in the transverse direction. 
All the other degenerate states of the planar phase are obtained by spin, orbital and phase rotations of the group $G= SO(3)_S \times  SO(3)_L \times U(1)$.
The Bogoliubov-de Gennes (BdG) Hamiltonian for fermionic quasiparticles in the chosen state in 
Eq. (\ref{OP1}) is:
\begin{equation}
 H({\bf p})=
   \begin{pmatrix}\epsilon(p) &\hat \Delta
  \\
  \hat\Delta^\dagger& -\epsilon(p)
  \end{pmatrix}
  =\epsilon(p) \tau_3 + c_\perp\tau_1(\sigma_x p_x +\sigma_y p_y)i\sigma_y\,,
\label{H}
\end{equation}
where $\tau_i$ are the Pauli matrices in the particle-hole space,
and $\epsilon(p)$ is the particle spectrum in the normal state of $^3$He, which in the vicinity of the Fermi surface at $p=p_F$ is $\epsilon(p)=c_\parallel(p-p_F)$. Here $c_\parallel=v_F$ is the Fermi velocity of the normal Fermi liquid. In the planar phase $c_\parallel$ plays the role of the "speed of light" propagating along $z$.

For us it is important that the planar phase has the operator of discrete symmetry $C=\sigma_3$, which commutes with the above BdG Hamiltonian:\cite{Makhlin2014} 
\begin{equation}
CH=HC \,.
\label{C}
\end{equation}
$C$ is the combined symmetry: the combination of spin rotation by the angle $\pi$ about the $z$-axis and a rotation of the phase of the BdG wave function by $-\pi/2$. In terms of the order parameter  $A_{\alpha}^i$ in Eq. (\ref{OP1}) it is the spin rotation by $\pi$ followed by a change of the phase of the order parameter by $\pi$. 

There is also the other representation of the planar phase, which is frequently used:
\begin{equation}
\tilde H =U^\dagger H U\,\,,\,\, U= \left( \begin{array}{cc}
1 & 0 \\
0 &i\sigma
 \end{array}
 \right).
 \label{Presentation}
\end{equation}
In this representation the Hamiltonian and the corresponding symmetry operator are:
\begin{equation}
 \tilde H=\epsilon \tau_3 + c_\perp\tau_1(\sigma_x p_x +\sigma_y p_y) \,\,, \,\, \tilde C=\tau_3\sigma_z
 \,\,, \,\, \tilde C \tilde H= \tilde H \tilde C\,.
\label{Crotated}
\end{equation}

\section{Spinor coordinate in the planar phase}
\label{SpinorPlanar}

The reason why we considered the symmetry $C$  is that it demonstrates how the discrete symmetry of the Hamiltonian can be extended to continuous $U(1)$ symmetry of the same Hamiltonian. The single-particle Hamiltonian in Eq.(\ref{H}) commutes  with the transformations $U_C(\alpha)$ generated  by $C$:\cite{Makhlin2014}
 \begin{equation}
U_C(\alpha)= \exp\left({-i\frac{\alpha}{2}}\right)\left( \cos \frac{\alpha}{2} + i C \sin \frac{\alpha}{2}\right) \,\,, \,\, C=U_C(\pi)\,.
\label{CExtended}
\end{equation}
Here the parameter $\alpha$, with $0\leq \alpha \leq 2\pi$, corresponds to the new spinor coordinate.
For $\alpha=\pi$ this is our discrete  symmetry $C$, i.e. $U_C(\alpha=\pi)=C$. In general, the operator $U_C(\alpha)$ describes the spin rotation by angle $\alpha$ followed by rotation of the phase 
$U(-\alpha/2)$ (since we discuss the pair condensate, $U(-\alpha/2)$ corresponds to rotation by $-\alpha$ of the phase of the order parameter).
This extension of discrete symmetry $C$ to continuous symmetry $U_C(\alpha)$ represents the extra (hidden) symmetry of the planar phase.  The same can be done with the planar phase Hamiltonian in the representation of Eq. (\ref{Presentation}). In this case  the discrete symmetry $\tilde C$ is extended to the continuous symmetry $U_{\tilde C}(\alpha)$.

The important property of such combined transformation of spin is the following. While the original $2\pi$  rotation of spin, $O(2\pi)$, changes the sign of the fermionic wave function, the new $2\pi$  rotation does not change the sign, since $U_C(2\pi)=O(2\pi) U(\pi) =1$. This is one of the possible realizations of the Andreev scenario. We introduced new spin degree of freedom -- the rotation angle $\alpha$ -- and obtained that both fermionic and bosonic wave functions are invariant under the $2\pi$ rotations.   
That is why in the Universe, in which such combined symmetry is operating, the fermion-boson transmutation is allowed. 

Of course, this does not solve the superselection problem in our Universe, because such compensation of the sign by extra spinor degrees of freedom  takes place only in a very special model --  the planar phase of superfluid $^3$He, and thus is not fundamental.  However, this example demonstrates that if something similar takes place on the trans-Planckian level, this would destroy the superselection conjecture. That is why the introduction of the additional degree in the internal space suggested by Andreev can be productive.

\section{Spinor coordinate for Standard Model extension}
\label{SpinorSM}

Interestingly,  a similar extension of the discrete symmetry to the continuous symmetry has been suggested for Dirac spinors in the Standard Model.\cite{Maiezza2022}
Let us consider the parity transformation. It is the combined symmetry $P=P_cP_s$, which contains the coordinate transformation  $P_c\Psi({\bf r},t)=\Psi(-{\bf r},t)$ and the internal symmetry of spinor $P_s=\gamma_0$, where $\gamma_a$ with $a=(0,1,2,3)$ are the Dirac matrices.
Now similar to Eq. (\ref{CExtended}), we can introduce the  following extension of the discrete internal parity $P_s$ of spinors to the combined continuous symmetry:
 \begin{equation}
U_{P_s}(\alpha)= \exp\left({-i\frac{\alpha}{2}}\right)\left( \cos \frac{\alpha}{2} + i P_s \sin \frac{\alpha}{2}\right) \,\,, \,\, P_s=U_{P_s}(\pi)\,.
\label{PExtended}
\end{equation}

So, in addition to the Lorentzian spin group $SO(1,3)$ with 6 generators we have one extra generator for internal degrees of freedom. As is shown in Ref.\cite{Maiezza2022}, such an extension can be naturally included in the $SO(1, 4)$ group.
As a result, one has the unusual situation, when the coordinate space is (3+1)-dimensional and obeys the  $SO(1,3)$ group with 6 generators, while the internal spin space is (4+1)-dimensional and obeys the  $SO(1,4)$ symmetry group with 10 generators. Instead of one extra degree of freedom in the planar phase, in the relativistic theory we have $10-6=4$ extra spin degrees of freedom.

When gravity is considered, such extension generates the gravity in terms of the rectangular (non-quadratic) vielbein,\cite{Volovik2020,Volovik2022b} see Sec.\ref{Rectangular} below.

\section{Spinor coordinates and rectangular vielbein:}
\label{Rectangular}

Let us consider the relativistic theory in which the dimension of the internal spin space is larger than dimension of the coordinate space. For the internal  $SO(1, 4)$ group of the spin space we need five anticommuting matrices $\Gamma^a$ with $a=(0,1,2,3,4)$. They can be constructed from the Dirac $\gamma$-matrices:
 \begin{eqnarray}
\Gamma^0 =\gamma_0= \tau_1 \,\,,\,\, a=0
\label{GammaMatrices1}
\\
 \Gamma^a =\gamma_a= i\tau_2 \sigma^a \,\,,\,\, a=1,2,3 \,,
\label{GammaMatrices2}
\\
 \Gamma^4 = -i \gamma_5= i\tau_3 \,\,,\,\, a=4 \,.
\label{GammaMatrices3}
\end{eqnarray}
The extra 4 generators of the $SO(1, 4)$ symmetry group are $\sigma^a \tau_1$, where $a=1,2,3$, and $i\tau_2$. One can check, that each of the three operators $\sigma^a \tau_1$ may play the role of the internal parity $P_s$ in Eq. (\ref{PExtended}), i.e. they represent different extensions of the discrete parity to continuous symmetries.\cite{Maiezza2022} let us now consider, how the extra internal spin dimension influence the gravity.

Conventional tetrad gravity in the Einstein–Cartan–Sciama–Kibble theory is described in terms of the quadratic $4\times 4$ tetrads.
The inverse Green's function of Dirac fermions  is expressed in terms of the tetrad (vierbein) $e^\mu_a$ in the following way:
 \begin{eqnarray}
G^{-1}(M) = e^\mu_a\Gamma^a p_\mu +M \,,
\label{4times5}
\end{eqnarray}
where the spin and coordinate indices of the vielbein $e^\mu_a$ are $a=(0,1,2,3)$ and $\mu =(0,1,2,3)$.
In the theory with extended spin space, instead of the quadratic $4\times 4$ gravitational tetrads, we have the $4\times 5$ rectangular gravitational vielbein. This is because the coordinate space is the conventional (3+1)-dimensional spacetime, while the spin space has the dimension (4+1). The corresponding inverse Green's function of Dirac fermions  is expressed in terms of the rectangular vielbein $e^\mu_a$ in the same way as in Eq.(\ref{4times5}), but now the spin and coordinate indices of the vielbein $e^\mu_a$ are $a=(0,1,2,3,4)$ and $\mu =(0,1,2,3)$.

Although the viellbein is rectangular the coordinate spacetime is still described by the  $4\times 4$ metric. This can be seen from the spectrum of Dirac fermions, which can be obtained from the product of two matrix Green's functions with opposite signs of the mass terms:
 \begin{eqnarray}
G^{-1}(-M) G^{-1}(M) = (e^\mu_a\Gamma^a p_\mu  +M)( e^\nu_b\Gamma^b p_\nu  -M)=
 -(\eta^{ab}e^\mu_a e^\nu_b p_\mu p_\nu +M^2)\,,
\label{gmunu2}
\end{eqnarray}
where
 \begin{eqnarray}
\eta^{ab}=(-1,1,1,1,1) \,.
\label{gmunu3}
\end{eqnarray}
This shows that the poles of the Green's function describe the spectrum of massive Dirac particles 
 \begin{eqnarray}
g^{\mu\nu}p_\mu p_\nu +M^2=0\,,
\label{Spectrum}
\end{eqnarray}
in the (3+1)-dimensional spacetime with the metric $g^{\mu\nu}$:
 \begin{eqnarray}
g^{\mu\nu} =\eta^{ab}e^\mu_a e^\nu_b\,.
\label{gmunu3}
\end{eqnarray}

The extra dimensions of spin space are hidden inside the metric. That is why, if only the metric of spacetime is known one has no complete  information on the spin degrees of freedom in our quantum vacuum. On the level of the metric one cannot resolve between different dimensions of the internal space and its transformation properties are not known. This justifies  the concern by Andreev, that the symmetry arguments in support of the superselection rule may not work. The hidden spin degrees of freedom may destroy the superselection conjecture.

\section{Rectangular vielbein in planar phase and its hidden gravitational global monopole}
\label{VielbeinPlanar}

In the planar phase of superfluid  $^3$He the relativistic physics emerges in the vicinity of the two Dirac points in the quasiparticle spectrum at ${\bf p}=\pm p_F \hat{\bf z}$. 
If the $s$-wave component is added  to the planar phase order parameter, the mass $M$ of the Dirac fermions is produced.\cite{Volovik2022b}
It appears that such relativistic physics experiences an extension of spin space similar to the Standard Model scenario in Sec. \ref{SpinorSM}.   The relativistic physics in the planar phase is described by the vielbein with mixed dimensions (4+1 dimension of internal space  and 3+1 dimension of spacetime coordinates).\cite{Volovik2020,Volovik2022b}

In the $^3$He notations, the Dirac $\Gamma$-matrices in Eqs.(\ref{GammaMatrices1})-(\ref{GammaMatrices3}) have the following form:
\begin{equation}
\Gamma^0=i\tau_2 \,\,, \,\Gamma^a=\tau_3\sigma^a \,\, (a=1,2,3)\,\,, \,    \Gamma^4=\tau_1 \,.
\label{5gamma}
\end{equation}
For the general order parameter in the planar phase, the Eq.(\ref{Crotated}) for BdG Hamiltonian has the following form near the two Dirac points at ${\bf p}=\pm p_F\hat {\bf l}$, where $\hat {\bf l}$ is the orbital unit vector:
\begin{equation}
  \tilde H=\sum_a \Gamma^a e_a^i (p_i - qA_i)  \,.
\label{HamiltonianPlanarRel}
\end{equation}
 Here ${\bf A}=p_F\hat {\bf l}$ is the vector potential of the effective gauge field acting on the massless Dirac fermions; $q=\pm 1$ is the corresponding effective electric charge. Since Eq. (\ref{HamiltonianPlanarRel}) describes the Hamiltonian, the index $a$ in 
the $\Gamma^a$ matrices  has the values $a=1,2,3,4$, and the $3\times 4$ matrix $e_a^i$ contains the components of the spatial vielbein  with $a=1,2,3,4$ and $i=1,2,3$:\cite{Volovik2022b}
\begin{equation}
 e^i_a= c_\perp (\delta_a^i -\hat l_a \hat l^i )\,\,\,  {\rm for} \,\, a=1,2,3\,\,, \, e^i_4=c_\parallel \hat l^i\,.
\label{vielbein}
\end{equation}
For simplicity, we ignore here the mixed components $e^0_a$ and $e^i_0$.

From the rectangular $3\times 4$ vielbein one obtains the conventional $3\times 3$ spatial metric:
  \begin{equation}
g^{ik}=\sum_{a,b}\delta^{ab} e^i_a  e^k_b \,\,, \,  a,b=1,2,3,4  \,\,, \,  i,k=1,2,3 \,,
\label{3Dmetric1}
\end{equation}
which has the conventional form
  \begin{equation}
g^{ik}= c_\parallel^2 \hat l^i  \hat l^k + c_\perp^2 (\delta^{ik}-\hat l^i  \hat l^k)\,.
\label{3Dmetric2}
\end{equation}
Again, the metric has no information on the hidden extra spin degrees of freedom, which may support the Andreev's hypothesis of the fermion-boson transmutation.

The interesting consequence of the extension of the spin degrees of freedom is the unusual structure of 
the topological object --  the monopole in the vielbein field. The simplest example of this gravitational monopole is the hedgehog with $\hat {\bf l}({\bf r})= \hat {\bf r}$. In the chiral superfluid $^3$He-A, such monopole terminates the quantized vortices of different types. It is the analog of the Nambu monopole terminating cosmic string—the observable Dirac string. In the planar phase, the monopole
is topologically stable and the Nambu–Dirac string does not appear. This is the result of the specific  spin degrees of freedom.\cite{Volovik2020}

When $c_\parallel^2 =c_\perp^2\equiv c^2$,  the metric in Eq. (\ref{3Dmetric2}) becomes flat and corresponds to the Minkowski vacuum in general relativity:
 \begin{equation}
g^{ik}=c^2\delta^{ik}\,.
\label{3Dmetric2}
\end{equation}
However, the topological monopole structure of the vielbein field does not disappear. This hedgehog is hidden from the observers who measure only the gravitational field.
This object with the hidden spin degrees of freedom is very different from the conventional global monopole, which generates  the solid angle deficit in the spacetime outside the monopole.\cite{Vilenkin1989} This is the unexpected consequence of the Andreev proposal for the hidden spin degrees of freedom.

 \section{Superfluids vs time crystals}
 \label{TimeCrystal}

The problem of transmutations between fermions and bosons is closely related to the 
problem of the existence of system with variable number of fermions. That is why most of Andreev's arguments were based on the consideration of the coherent superposition of states with different conserved charges (angular momentum, particle number, etc.).
The coherent superposition of such states is time dependent. That is why, according to a paper by Andreev from 1996 \cite{Andreev1996}, the spontaneous breaking of the global $U(1)$ symmetry accompanying the Bose condensation corresponds to thermodynamically equilibrium ground states with non-integral average particle number. The latter results in the time-dependent order parameter: 
\begin{equation}
\left<\hat a^+ \right>= {\cal N}^{1/2} \exp{\left( i\frac{\mu}{\hbar}t\right)}\,,
\label{oscillations}
\end{equation}
where $\hat a^+$ is the creation operator of bosons, $\mu$ is their chemical potential, and $ {\cal N}$ is the number of particles in Bose condensate. Eq. (\ref{oscillations}) suggests that  there are oscillations with frequency $2\pi \hbar/\mu$, and thus this looks like a spontaneous breaking of the symmetry under the time translation.

The problem is that such oscillations are produced by the system in its ground state, which is rather strange. However,  Wilczek in 2012 introduced the notion of the time crystals,\cite{Wilczek2012} which in particular included the Bose condensates (see e.g. the papers 
"Superfluidity and space-time translation symmetry breaking"\cite{Wilczek2013} and "Space- and time-crystallization effects in multicomponent superfluids"\cite{Prokofev2020}).
Although such realizations of time crystals were criticized,\cite{Bruno2013a,Bruno2013b,Nozieres2013}  time crystals are now a hot topic.  

Clearly, a system in its ground state cannot have physically observable oscillations. If the oscillations can be detected using some stationary device, this would mean that the detector is excited. The energy of the system is transferred to the detector, which means that the system is not in the ground state. So, if the Bose condensate is in its stable ground state, such as superfluid liquid $^4$He with $\mu<0$ at $T=0$, the oscillations of the order parameter in Eq. (\ref{oscillations}) are not observable. 

In principle, there are several ways the oscillations can be observed, but in all cases the system is either perturbed, or the system is interacting with environment, or the number of particles is not strongly conserved. The latter takes place if we consider the possibility of proton decay suggested by the Grand Unification Theories (GUTs), which in particular  leads to the non-conservation of the helium atoms.
Measuring the intensity of the decay of helium atoms in superfluid $^4$He, one can observe oscillations between states with different numbers of atoms in the condensate.

Another source of the decay of superfluid $^4$He is the expansion of the Universe. The probability of evaporation of atoms from this Bose liquid is:\cite{Volovik2009}
\begin{equation}
w \propto \exp{\left( -\frac{\pi|\mu|}{\hbar H}\right)}\,.
\label{decay}
\end{equation}
where $H$ is the Hubble parameter. This looks as if the quantum vacuum in the expanding Universe has the effective temperature $T=H/\pi$, which leads to the activation process of evaporation  of atoms from the liquid to the vacuum. Note that this decay is the local process, which is not related to the cosmological horizon. The corresponding temperature, which determines the thermal-like activation, is twice the Gibbons-Hawking temperature, see also Ref. [\cite{Volovik2023}].

So, in principle, the oscillations are observable, but they become unobservable in the limit of the infinite decay time, when the system can be considered in its ground state. That is why the notion of time crystals requires consideration of the interplay between different time scales.  The typical example is provided by the spontaneous coherent precession of spin, which can be considered in terms of the Bose condensate of magnons.\cite{Volovik2013,BunkovVolovik2013} The corresponding $U(1)$ symmetry is the symmetry under spin rotations around the direction of the magnetic field and the corresponding $U(1)$ charge is the spin projection on the direction of the magnetic field or, which is the same,  the number of magnons.

Both the coherent spin precession and the Bose condensates in superfluids experience  the off-diagonal long-range order. This is seen if one compares the operator of creation of particle creation 
$\hat a_0^+$  with  the operator $\hat S^+$ for the creation of spin projections on the $z$ axis, whose expectation value  is:
\begin{equation}
\left<\hat S^+ \right>={\mathcal S}_x +i {\mathcal S}_y =S_\perp e^{i\omega t}\,,
\label{spinODLRO}
\end{equation}
where $\omega$ is the precession frequency.

We can use the Holstein-Primakoff transformation, which expresses the spin operators in terms of the operators of creation and annihilation of magnons:
\begin{eqnarray}
\hat a_0~\sqrt{1-\frac{\hbar a^\dagger_0 a_0}{2{\mathcal S}}}= \frac{\hat {\mathcal S}^+}{\sqrt{2{\mathcal S}\hbar}}~,
\label{MagnonAnnih}
\\
\sqrt{1-\frac{\hbar a^\dagger_0 a_0}{2{\mathcal S}}}~\hat a^\dagger_0=
\frac{\hat {\mathcal S}^-}{\sqrt{2{\mathcal S}\hbar}}~,
\label{MagnonCreation}
\\
\hat {\mathcal N}=\hat a^\dagger_0\hat a_0 =
\frac{{\mathcal S}-\hat{\mathcal S}_z}{\hbar}=
\frac{{\mathcal S}(1-\cos\beta)}{\hbar} \,,
\label{MagnonNumberOperator}
\end{eqnarray}
where $\beta$ is the tipping angle of spin precession.
This gives Eq. (\ref{oscillations}) for the  vacuum expectation value  of the operator of boson annihilation 
\begin{equation}
\left<\hat a_0 \right>={\mathcal N}^{1/2}  e^{i\omega
t}=\sqrt{\frac {2{\mathcal S}}{\hbar}} ~\sin\frac{\beta}{2} ~e^{i\omega
t}\,,
\label{ODLROmagnon}
\end{equation}
which shows that the precession frequency plays the role of  the chemical potential of magnons,
$\hbar\omega=\mu$, in agreement with Eq.(\ref{oscillations}). 

This time dependence is observable: the spin precession is seen in the NMR experiments due to the spin-orbit interaction. On the other hand, the spin-orbit interaction violates the $U(1)$ symmetry of spin rotations, which leads to the non-conservation of the number ${\mathcal N}$ of magnons in the magnon Bose condensate. If the spin-orbit interaction is neglected, the number of magnons is conserved, and the coherent spin precession represents a time crystal -- it is the ground state of the system at a fixed number ${\mathcal N}$ of magnons. But without the spin-orbit coupling the precession is not observable,  the oscillations would be hidden and the detector would not detect the time crystal  (we are again faced with hidden spin degrees of freedom). 

This is the essence of the physical time crystals. It applies also to all other coherent systems generated by the quasi-conservation of the $U(1)$ charge $Q$. Each system is characterized by its own relaxation time $\tau_Q$, which is the decay time of the corresponding charge. The coherent state is time dependent, i.e. it violates the time translation symmetry. But it is not the ground state of the system, since the ground state of the decaying system has zero charge, $Q=0$. In the strict limit 
$\tau_Q \rightarrow \infty$ the coherent state becomes the ground state at fixed charge $Q$. But in this limit  the oscillations become un-observable: no breaking of time translation symmetry is seen in this limit.

So in general,  in systems with quasi-conserved charge $Q$ the time crystals do exist and they are observable. For example, due to the long relaxation time $\tau_Q$ in $^3$He-B, we observe the AC Josephson effect between two time crystals.\cite{Autti2021} This demonstrates that the spontaneous breaking of time translation symmetry suggested by Andreev does take place.

Andreev also applied this symmetry-breaking notion to fermions, and not only to the pair condensate, which looks natural, but also to superpositions between states with odd and even numbers of fermions.\cite{Andreev2003} This remains controversial, but it is not excluded that something like that can occur due to the quantum effects related to the chiral anomaly, which we consider in the next section \ref{Anomaly}.

\section{Chiral anomaly and non-conservation of fermion number}
\label{Anomaly}

It is well known that the violation of fermion number conservation law is possible in gauge theories with nontrivial topology. In the instanton process of the transition between the quantum vacua with different topological charges, the baryon and lepton numbers  ${\cal N}$ are not conserved. This is the basis of the Kuzmin–Rubakov–Shaposhnikov scenario of the anomalous electroweak baryogenesis.\cite{Rubakov1985}
 In the Standard Model the instanton process leads to non-conservation of  ${\cal N}$
 by an even number, while Andreev advocated the possibility of  non-conservation  
 by an odd number of fermions.
 
 The same non-conservation  by an even number of fermionic quasiparticles takes place in condensed matter systems. For example, in superfluid $^3$He-A  the gravitational anomaly leads to the creation of an even number of quasiparticles. The gravitational instanton process of creation of a single hopfion is accompanied by the creation
of 6 chiral fermions:\cite{Volovik2021} 
\begin{equation}
  \partial_\mu n^\mu_H = 6\,\partial_\mu J^\mu_5\,.
 \label{GravitationalAnomalyInstanton}
\end{equation}
Here $J^\mu_5$ is the chiral current of fermionic quasiparticles in  $^3$He-A , and $n^\mu$ is the hopfion current, with the hopfion density
\begin{equation}
n^0_H= \frac{m^2}{4\pi^2}   ({\bf v}_s\cdot(\nabla\times{\bf v}_s)) \,\,,\, 
N_H=  \int d^3r \,n^0_H\,,
 \label{HopfionCharge}
\end{equation}
where ${\bf v}_s$ is superfluid velocity and $m$ is the mass of the $^3$He atom.
This is the gravitational analogue of the Kuzmin–Rubakov–Shaposhnikov scenario of the
anomalous electroweak baryogenesis.\cite{Rubakov1985}

Some gauge theory models allow topological processes that lead to the creation
of just one fermion. These theories contradict different principles of quantum field theory, including the spin-statistics theorem. And usually, there exist arguments as to why such models cannot be realized.

However, following Andreev's arguments, Bezrukov, Burnier and Shaposhnikov\cite{Shaposhnikov2006} found that at least in 1+1 dimensions it is possible to create a single  fermion (see also Ref.\cite{Burnier2006}). It is created  in the
instanton process, in which the topological charge of the quantum vacuum changes by one.
This shows that it is in principle possible that Nature may have some kind of hidden channel that allows the fermion-boson transformation. An extra spin dimension in trans-Planckian physics may be a possible route to such processes.

 \section{Non-conservation of fermion number by Andreev-Weyl fermions}
 \label{AndreevWeyl}
 
Here we discuss another possible extension of relativistic quantum field theory, which allows the creation of a single  fermion. Let us consider two Weyl points separated in momentum space:
\begin{equation}
 H({\bf p})=e^i_\alpha \sigma^\alpha (p_i - p^0_i)+f^i_\alpha \sigma^\alpha (p_i +p^0_i) \,.
  \label{TwoWeyl}
\end{equation}
This Hamiltonian describes two different "worlds" of Weyl. fermions with two different tetrad fields. One is the world of the Weyl particles concentrated near the Weyl point at ${\bf p}={\bf p}^0$ with the gravitational tetrads $e^i_\alpha$, while the other particles live near the Weyl point at  ${\bf p}=-{\bf p}^0$ with the gravitational tetrads $f^i_\alpha$.
If the distance $2p^0$ between the Weyl points is on the order of the Planck scale, these worlds practically do not communicate with each other. Each world has its own gravity with its own metric. In each world the scattering of a particle practically does not change its momentum, i.e.  ${\bf p} ={\bf p}^0 + {\bf k} \rightarrow {\bf p} ={\bf p}^0 - {\bf k}$ with $k\ll p^0$. This is the analogue of Andreev reflection, where the  momentum practically does not change but the velocity $dE/d{\bf p} \sim {\bf k}/M$ changes sign. That is why we can call such particles Andreev-Weyl fermions.

Nevertheless, there exists  the true reflection, where the momentum changes sign, ${\bf p} \rightarrow -{\bf p}$. The probability of such processes is extremely small, since it is not easy to create a perturbation on the Planck energy scale, which can lead to a reflection, such as as mirror of Planck length width.  Nevertheless, such a reflection is possible, and then in each world, the observer will detect the creation or annihilation of a single fermion. In this case,  for the world near the first Weyl points, the role of the hidden spin variable is played by the tetrad $f^i_\alpha$, which is hidden in the second world.

Models of the type of Eq. (\ref{TwoWeyl}) are known for lattice fermions.\cite{NielsenNinomiya1981a,NielsenNinomiya1981b}
The natural arrangement of the eight left and eight right Weyl fermions in one generation of the Standard Model is when the Weyl points are on the vertices of the cube in the four-dimensional momentum-frequency space.\cite{Creutz2008,Creutz2014} The  analogous  three-dimensional cube of the Weyl points in $d$-wave superconductors and some other possible configurations of Weyl and Dirac points are discussed in Ref. [\cite{Volovik2017}], which is devoted to the memory of Lev Petrovich Gor’kov.

\section{Discussion}

While the specific scenario suggested by Andreev is not realized, his suggestion of extra dimensions for the spin space and the corresponding non-conservation of fermion number may occur in different realizations of the trans-Planckian physics. We considered several possible scenarios: 
(a) extension of the discrete parity to the continuous spin symmetry group on example of the planar phase of superfluid $^3$He; (b) extension of the internal spin symmetry group $SO(1,3)$ to $SO(1,4)$ in the relativistic system; (c) a similar extension in relativistic physics emerging in the vicinity of the Dirac points in the planar phase; (d)  extension of the Standard Model, which allows the baryogenesis; (e) the 1+1 quantum field theory, which allows the fermion-boson transmutation; (f) extension of Standard Model to lattice theory.

In case (a) the extra spin dimension compensates for the change of sign of the fermionic wave function under $2\pi$ rotation. This demonstrates, that the Andreev-type scenario, which destroys the superselection rule, may in principle take place on a more fundamental level of trans-Planckian physics. In this case we must look for a different solution to the fermion-boson problem.

In cases (b) and (c) the extra spin dimensions lead to new physical effects. The extra internal degrees of freedom give rise to gravity based on a rectangular vielbein rather than on traditional tetrads of Einstein-Cartan gravity. It is interesting that these extra spin dimensions remain hidden, if one only probes the metric in experiments. An example is the "invisible" gravitational global monopole in the planar phase with isotropic "speed of light" in Sec. \ref{VielbeinPlanar}. It is the singular topological structure (hedgehog) in the vielbein field, which however does not disturb the Minkowski metric. 

In case (d) the baryogenesis leads to the decay of atoms in superfluid $^4$He, which in turn produces the time crystal behavior of superfluids advocated by Andreev. 

Case (e) is based on consideration of Andreev's arguments. It is shown that in the 1+1 theory  the anomalous creation of single fermions is possible. This opens the route to the anomalous fermion-boson transmutation using extra dimensions.

In case (f) the extra degrees of freedom correspond to different worlds emerging close to the widely separated Weyl points in  momentum space. Within each of these separate worlds the observer may detect events of single fermion creation, while the other fermions are hidden in the other worlds. 

So, the Andreev's idea of extending the internal spin space, even in case if it does not destroy the superselection rule, may lead to new physics on a more fundamental level.

 {\bf Acknowledgements}.   
 This work has been supported by the Academy of Finland (grant 332964).

\end{document}